\documentclass[12pt]{article}
\usepackage{epsfig}
\usepackage{amssymb}
\usepackage{pslatex}
\usepackage{array}
\usepackage{calc}

\makeatletter

\@addtoreset{equation}{section}
\makeatother
\renewcommand{\arraystretch}{1.24}

\newcommand{\bea}{\begin{eqnarray}}
\newcommand{\eea}{\end{eqnarray}}
\newcommand{\be}{\begin{eqnarray}}
\newcommand{\ee}{\end{eqnarray}}

\begin{document}

\begin{titlepage}
\vskip1cm
\begin{flushright}
$\mathbb{UOSTP}$ {\tt 10301}
\end{flushright}
\vskip1.25cm
\centerline{\large \bf
Static Length Scales of ${\cal N}=6$ Chern-Simons Plasma
}
\vskip1.25cm
\centerline{  Dongsu Bak$^a$, Kazem Bitaghsir Fadafan$^b$, Hyunsoo
Min$^a$
}
\vspace{1.25cm}
\centerline{\sl a) Physics Department, University of Seoul, Seoul
130-743 {\rm KOREA}}
\centerline{\tt \small (\,dsbak@uos.ac.kr, hsmin@dirac.uos.ac.kr\,) }
\vskip0.25cm
\centerline{\sl b) Physics Department,
Shahrood University of Technology}
\centerline{\sl  P.O.Box 3619995161, Shahrood {\rm IRAN}}
\centerline{\tt \small (\,bitaghsir@shahroodut.ac.ir\,) }
\vspace{2cm}
\centerline{ABSTRACT}
\vspace{0.75cm}
\noindent
Using gravity description, we 
compute various static length scales of 
 ${\cal N}=6$ Chern Simons plasma
in a strongly coupled regime.  For this, we consider the $\mathbb{CP}_3$ 
compactification of the type IIA supergravity 
down to four dimensions, and identify all the low-lying bosonic modes
up to masses corresponding to the operator dimension $3$ 
together with all the remaining
$\mathbb{CP}_3$ invariant modes. 
We find  the true mass gap, the Debye screening 
mass and the corresponding dual  operators 
to be probed in the field theory side.

\vspace{1.5cm}
\end{titlepage}

\section{Introduction}
There has been great progress in 
understanding of AdS/CFT correspondence between 
string theories and gauge theories\cite{Maldacena:1997re}. 
In addition to the well known duality between
the type IIB string theory on AdS$_5\times S^5$ background and  ${\cal N}=4$
super Yang-Mills (SYM) theory\cite{Maldacena:1997re}, 
the type IIA version of AdS$_4$/CFT$_3$ correspondence 
is put forwarded recently.
The theory now known as the ABJM model is the three dimensional ${\cal N}=6$
U(N)$\times\overline{{\rm U}}$(N) superconformal Chern-Simons theory with level
(\,$k$,$-k$\,) and proposed to be dual to the type 
IIA string theory on AdS$_4 \times\mathbb{CP}_3$
background\cite{ABJM}.  
Some test of this duality has been carried out based on the 
integrabilities with
indication of some additional interesting 
structure\cite{Minahan:2008hf,Bak:2008cp,Bak:2009mq}. 

The type IIA supergravity description is dual to the large 
$N$ planar limit  of the Chern-Simons
theory where one is taking $N,\,\, k \rightarrow \infty$  
while holding the 't Hooft coupling
$\lambda ={N\over k}$ fixed. Some probes of response 
properties of  the finite temperature
Chern-Simons theory are studied recently via the 
$\mathbb{CP}_3$ invariant dimensional 
reduction of the type IIA supergravity\cite{Yun}. 
For the study of its thermodynamic properties 
in the weak coupling regime, see Ref.~\cite{Smedback:2010ji}.
  
In this note, we shall  be concerned about the physics of the  
finite temperature
ABJM plasma at strongly coupled regime and study its various static 
length scales.
These   are arising as spatially decaying 
scales when local operators 
are inserted at a certain point of the finite temperature plasma.
Among them, we are particularly interested in the 
two scales of   true mass gap $m_g$ and 
Debye screening mass $m_D$. The inverse of the longest 
possible correlation length scale
is the definition of the true mass gap while the lowest in the $CT$ 
odd sector\footnote{Unlike usual Chern-Simons theories,
the ABJM theory has the parity $P$ and the time-reversal $T$ 
symmetries. The standard parity and time-reversal transformations of the
ABJM theory  bring
its level $(k,-k)$ to $(-k,\,k)$. To define our $P$ and $T$, one further
interchanges the U(N) and the $\overline{\rm U}$(N) gauge fields
together with yet further exchange of all the matter fields 
with their adjoints, 
which brings its level back to $(k,-k)$. The charge conjugation $C$ 
is defined in the standard manner.}
is called the Debye screening mass following the definition of  
Yang-Mills plasmas\cite{Arnold:1995bh,Bak:2007fk}. For the
${\cal N}=4$ SYM theory, these scales are identified at strong 
coupled regime in Ref~\cite{Bak:2007fk}. 
It is argued there that these scales 
are well representing the universal characteristics of various 
Yang-Mills  theories. 
For instance the ratios $m_D/m_g$
for the ${\cal N}=4$ SYM theory and the $N_f=2$ QCD are 
matching with each other
in the strong coupling limit, 
which supports  such a picture\cite{Bak:2007fk}.

In this note, we shall compute these length scales of the ABJM 
plasma using the 
description of  black brane background, which has
validity for the strongly coupled regime.
For this purpose, we consider the $\mathbb{CP}_3$ compactification 
of the type IIA supergravity
down to four dimensions\cite{Nilsson:1984bj}
 and identify all the low lying bosonic modes up to 
masses corresponding to operator dimension $\Delta =3$ 
together with all the remaining
 $\mathbb{CP}_3$ invariant higher dimensional 
modes\cite{Yun}.
They are described by 4d scalar, 
vector and graviton equations in the bulk of the black brane background.
 For each gravity mode, one sets up eigenvalue problem with specific 
boundary conditions\cite{Brower:2000rp} 
and find  mass (1/length) scales from 
eigenvalues determined by numerical analysis.
We then identify the true mass gap, the Debye mass and their dual operator 
contents.
 Dynamic 
(non static) responses of the finite temperature
system  are characterized by
equilibration time scales\cite{Bak:2007qw} 
and study of ABJM theory in this respect 
will be of interest. Also study of  static scales in 
the $U(1)$ charged  plasma of the ABJM theory\cite{Yun}
will be interesting in many respect\cite{Rey:2008zz}.   

In Section 2, we introduce the field theory definition 
of  static length (1/mass) scales in terms of 
two point correlator of field theory operator.
In Section 3, we consider the $\mathbb{CP}_3$ compactification and identify
the low-lying supergravity modes.  In Section 4, we  find a 
linearized fluctuation 
equation for  each mode, from which we determine 
length scales by  numerical analysis.
The results are summarized in Table 2. Last section is 
devoted to the discussion of the true
mass gap and the Debye mass. The relevant dual operator contents 
in the ${\cal N}=6$ Chern-Simons
theory are constructed  starting from  chiral primary operators
of the ABJM theory.


\section{Field theory definition of static length scales}

In this section, we are mainly interested in
static correlation length scales of the finite temperature
${\cal N}=6$ Chern-Simons plasma. From the view point of the field 
theory, one considers any gauge invariant operator $O({\vec r})$
which has an operator dimension $\Delta$. If this operator
is inserted at some point, {\it e.g.} ${\vec r}={\vec 0}$
of the finite temperature system, the 
perturbation is in general decaying exponentially
at large distance once the longest possible length scale 
is finite. This characteristic length scale is our 
concern in this note. More precisely they
can be measured by the study of
spatial behaviors of static two-point
correlator defined by
\be
{\cal C}(\vec{r})=\langle\, O^\dagger(\vec{r})\,\, O(\,\vec{0}\,)\,\rangle\,,
\ee
where the thermal expectation value is taken with respect to
the finite temperature vacuum state. 
We shall take the separation in the $x_1$ direction without loss
of generality.  Generic features of the above correlator
can be understood as follows. To define the finite temperature 
field theory, we use the Euclidean field theory where the 
Euclidean time direction is circle compactified
with size $\beta=1/T$. 
We then consider a Minkowski field 
theory where 
 the $x_1$ direction is Wick-rotated
from the Euclidean field theory.  
The fictitious time translation generator
is taken as our Hamiltonian $H_{x_1}$. Let us introduce
its eigenbasis by
\be
H_{x_1}\, |n\rangle =\epsilon_n\, |n\rangle\,.
\ee
Then the above correlator can be expanded as
\be
{\cal C}(x_1) =
\sum_n \, e^{-\epsilon_n\, x_1}\,\, |\,\langle n\,|\, O(\vec{0})\rangle\,|^2\,.
\ee 

At large distance, the lowest eigenvalue with nonvanishing 
amplitude $\langle n\,|\, O(\vec{0})\rangle$ basically controls the 
behavior of the two point correlator. We shall call the $i$-th lowest
eigenvalue with nonvanishing 
$\langle n\,|\, O(\vec{0})\rangle$ as $\mu_i$.
Projecting out the lowest nonvanishing contribution,
one may find the next decay mass scale $\mu_2$ 
and so on. Below we shall consider
various low dimensional operators and compute length scales
up to $\mu_3$ using the gravity 
description.

These scales of various operators with a definite dimension
are all important characteristics of the finite temperature
field theory itself. Among them, there are two particularly
interesting length scales. One is the longest possible
correlation length for all possible operators. The inverse of this scale
we call as the true mass gap $m_g$. The other scale
 is defined as
the longest possible length scale within the CT odd sector to which
the operators of charged excitation belong.
The inverse of this Debye length scale is called as
Debye mass $m_D$. 
This definition  agree with that of QCD or the ${\cal N}=4$ 
Super Yang-Mills 
theory\cite{Arnold:1995bh,Bak:2007fk}. 
These two are particularly important characteristic 
scales of  a gauge theory plasma.

Finding these two scales in the strongly coupled regime of 
${\cal N}=6$ Chern-Simons plasma will be the subject of 
the subsequent sections where we use 
the dual gravity description.
%
Among the low lying supergravity modes, we shall survey
possible eigenvalues of scales in order to find the true mass gap 
and the Debye mass. 

\section{$\mathbb{CP}_3$ compactification and the black brane background}

The IIA supergravity compactified on the internal 
$\mathbb{CP}_3$ space to the four dimensional spacetime has 
been much studied some time ago\cite{Nilsson:1984bj}. Its four 
dimensional spectra are all classified there, which we shall 
explain  to the extent we need for our purpose.
The bosonic spectra include towers of 4d scalars, vectors and spin 
two gravitons. These are classified by the symmetries
of U(1)$\,\,\times\,\,$SU(4) where SU(4) is the symmetry possessed
by the $\mathbb{CP}_3$ space and U(1) is from the
 circle related to the 11d interpretation.
The corresponding massless U(1) charge counting the D0 brane 
is interpreted as di-baryon charge of the ${\cal N}=6$ Chern-Simons 
theory\cite{ABJM}. For the SU(4) representation, we use the standard 
Dynkin labels $(l_1\, l_2\,l_3)$. The SU(4) singlet denoted by
$(0\,0\,0)$ is for the  $\mathbb{CP}_3$ invariant part of the 
spectra that form a consistent closed sector\cite{Yun}.

For the graviton modes we shall only consider the lowest mode
corresponding to the 4d AdS graviton fluctuation. There is of course
massive graviton tower which is not neutral under the SU(4) of 
$\mathbb{CP}_3$. 

For the spin one,
we shall consider the massless vector modes
of  $(0\, 0\, 0)$ and $(1\, 0\, 1)$ representations, where
$(1\, 0\, 1)$ corresponds to the adjoint representation of 
dimension $15$. In addition, we shall consider massive $M^2=2$
non-singlet gauge fields together with 
a singlet gauge field with $M^2=12$.

For scalars, the lowest modes consist of
$(1\, 0\, 1)^+$, $(1\, 0\, 1)^{-}$ and $(2\, 0\, 2)^{+}$ with 
$M^2=-2$.  We also consider non-singlet scalars with $M^2=0$ whose
representations are listed 
in Table 1. There are three singlet real scalars respectively
with $M^2= 4,10$ and $18$.  Therefore the whole singlet spectra
consist of one massless graviton, one massless and one massive vector
fields and 3 real scalars. Note that there is no fermionic 
singlet contributions at all. The full Lagrangian for this consistent
truncation of the IIA compactification is obtained in \cite{Yun}\footnote{
For the compactification from the M-theory view point, see also
Ref.~\cite{Gauntlett:2009zw}, in which the resulting spectra
involves extra modes not present in the type IIA 
compactification.}.    
The summary of the discussion is in Table 1.  
\begin{table}[ht]
{
\renewcommand{\arraystretch}{1.2}
\begin{tabular*}{150mm}{@{\extracolsep\fill}|l||l|l|l|}
\hline \hline
\phantom{aaaaaaaaaaaa}
& \phantom{abc}\phantom{aaa} spin 0 
& \phantom{abc} spin 1 
& \phantom{ab} spin 2 \phantom{abcd} \\
\hline \hline
$\Delta=1$ & $(101)^+_{-2}$ &   &   \\
\hline
$\Delta=2$ & 
$(202)^+_{-2}$ \ $(101)^-_{-2}$ 
 & $(000)^-_{0}$ \ $(101)^-_{0}$ &  \\
\hline
  $\Delta=3$ & $(303)^+_{0}$ \ $(202)^-_{0}$  $(400)^-_{0}$  & 
$(101)^-_{2}$ \ $(202)^-_{2}$ 
& \phantom{a} $(000)^+_{0}$\\
  &  $(004)^-_{0}$  $(210)^-_{0}$ $(012)^-_{0}$  
&  
$(210)^-_{2}$  $(012)^-_{2}$ 
  &  \\
  &   $(020)^-_{0}$   
&   &  \\
\hline 
${\bf 0}^{\rm th}$super\,\,multiplet 
&  $(101)^+_{-2}$\ $(101)^-_{-2}$ &  $(000)^-_{0}$\ $(101)^-_{0}$ 
&  \phantom{a} $(000)^+_{0}$   \\
\hline
{\small $\mathbb{CP}_3$} singlet sector 
&  $(000)^+_{4}$ $(000)^-_{10}$ $(000)^+_{18}$ 
&  $(000)^-_{0}$ $(000)^-_{12}$ 
&  \phantom{a} $(000)^+_{0}$      \\
\hline
\end{tabular*}
\caption
    {\small The low lying bosonic spectra up to the operator 
dimension 3 are presented. The upper and lower indices denote respectively
the parity and the mass squared value $M^2$. The bosonic part 
of the lowest ${\cal N}=6$ super multiplet and the whole spectra  
of $\mathbb{CP}_3$
singlet sector are presented in addition.}
\label{tab1} }
\end{table}

Turning off all the excited modes except the AdS gravity part, 
the 4d action becomes
\be
I_{4}= {1\over 16 \pi G_4}\int d^4 x \sqrt{-g}\Big(\,R+6\,\Big)
\label{effaction}
\ee
where the 4d Newton constant is given by\cite{Yun}
\be
{1\over 16 \pi G_4}= {N^2\over 12 \pi \sqrt{2\lambda}}\,.
\ee
Note also that we have scaled away the AdS curvature 
radius $R_s$ using the scaling property of the 4d action,
which is possible even including all the remaining modes.

We consider the well known black brane solution 
with a planar symmetry, in which the metric takes a form
\be
ds^2_4= {1\over z^2}\,\,\Big[\, h(z)\,dx_0^2 + dx_1^2+dx_2^2
+{dz^2\over h(z)} \Big]
\ee
with
\be
h(z)= 1- \left({z\over z_H}\right)^3\,.
\ee
This black brane background is dual to the finite 
temperature version of the ABJM theory. Due to the quantum scale 
invariance of theory, this finite temperature field theory 
depends on only one dimensionful parameter which is the 
temperature $T$. Hence the theory possesses only one finite-temperature
 phase 
corresponding to  high temperature limit.
This temperature 
is identified with the Hawking temperature of the black brane,
\be
T={1\over 4\pi}|h'(z_H)|={3\over 4\pi}{1\over  z_H}\,.
\ee
We further set the only length scale $z_H=1$ 
in the black brane metric 
as $h(z)=1-z^3$. This corresponds to the mass unit
${4\pi T}/3$, which can be recovered whenever needed.

\section{Static length scales}
In this section, we shall find the static mass scales
by analyzing  linear fluctuation equations in the black 
brane background. As explained in the previous section, there
are various 4d bulk modes up to spin 2, which are dual to 
field theory operators of definite scaling dimension.

We shall begin with the scalar mode whose equation takes the form
\be
\nabla^2 \Phi - M^2\, \Phi=0\,,
\ee
which is defined for the interval $z\in [0,1]$ for our Euclidean
black brane background. 
At the boundary $z=0$ where the field theory is defined, the scalar field
generally behaves as
\be
\Phi= a(x) \, z^{3-\Delta} + b(x)\, z^\Delta + \cdots
\ee
where $\cdots$ denotes all the remaining powers of $z$. The $\Delta$
here corresponds to the scaling dimension of the field theory
operator dual to $\Phi$. By a straightforward investigation, one finds
\be
\Delta ={1\over 2}\Big( 3+ \sqrt{9+ 4 M^2}\Big)\,,\phantom{\!\!=1}
\ee
with an exception of $\Delta=1$.
For this exception, one has
\be
\Delta ={1\over 2}\Big( 3- \sqrt{9+ 4 M^2}\Big)=1
\ee
with $M^2=-2$. 
To set up the eigenvalue problem, 
we consider the ansatz $\Phi= U(z)e^{-kx_1}$, 
which describes an exponentially decaying response of the perturbation.
The equation for $U(z)$ becomes
\be
z^4\,\Big(\,{h\over z^2}\,U'\Big)'+ \big(\, z^2 k^2 -M^2\,\big) U =0\,.
\label{scalareq}
\ee   
Having a second order equation,  one has in general two linearly 
independent solutions before imposing boundary condition
at the horizon. From the general analysis, the term $a(x)$ corresponds 
to adding the source term
\be
\delta\, I_{SCS}= \int d^3x \, \, a(x) \,\,O(x)
\ee
to the field theory and $b(x)$ is then interpreted as
the expectation value $\langle\,O(x)\,\rangle= b(x)$ 
where the expectation is taken in the presence of the source term
$a(x)$. For the scalar, we impose the boundary condition at the
horizon given by
\be
h(z) \, U'(z)|_{z=1}=0\,,
\ee
which kills the logarithmically singular term at the horizon.
This boundary condition insures that the on-shell supergravity 
action does not receive any boundary contribution at the 
horizon. In addition the finiteness of each individual term
in (\ref{scalareq}) is insured by the boundary condition.

For the eigenvalue problem, we shall further require the source term 
$a$ to vanish by adjusting $k^2$, which will fix the static 
scales of our concern. We consider the six lowest masses 
$M^2=-2,-2,0,4,10,18$ corresponding to operator dimensions
$\Delta=1,2,3,4,5,6$. The results are presented in Table 2 in the unit 
of $4\pi T/3$.

For massive vector modes, the Maxwell equation takes a form,
\be
\partial_\mu \big( \sqrt{g}\, F^{\mu\nu}\big)=M^2\sqrt{g}\,\big(\,
A^\nu+\nabla^\nu\varphi\big)\,,
\ee 
with
\be
\partial_\mu \,\Big[\sqrt{g}\,\big(\, 
A^\mu+\nabla^\mu\varphi\,\big)\,\Big]=0\,.
\ee
The scalar field $\varphi$ is arising as a 4d Poincar\`e dual to
three form field strength $H_{\mu\nu\lambda}$ and gets
absorbed into the massive gauge degree by the Higgs mechanism\cite{Yun}.
We use the $\varphi=0$ gauge leading to the condition,
\be
\partial_\mu \,\Big(\sqrt{g}\, 
A^\mu\,\Big)=0\,.
\ee
For the massless case of $M^2=0$, this condition does not follow 
from the equations of motion but can be incorporated as a part of 
the gauge fixing condition. Near boundary gauge fields in general
behave as
\be
A_{\bar\mu}= a_{\bar\mu}(x) \, z^{2-\Delta} + 
b_{\bar\mu}(x)\, z^{\Delta-1} + \cdots
\ee
where $\bar{\mu}=0,1,2$ is for the boundary directions at 
$z=0$ and the dimension of the dual current operator
is given by
\be
\Delta ={1\over 2}\Big( 3+ \sqrt{1+ 4 M^2}\Big)\,.
\ee 

For the length scales, we start with
an ansatz,
$A_\mu= {\cal A}_\mu (z)\, e^{-kx_1}$
and consider the massive case first. Turning on 
only $A_2$ leads to a consistent equation
\be
z^2\Big({h}\,{\cal A}'_2\Big)'+ \big(\,z^2 k^2  -M^2\,\big)\, {\cal A}_2=0\,.
\label{a1}
\ee
Similarly turning on only $A_0$ leads to
\be
z^2\,{h}\,{\cal A}''_0+ \big(\,z^2 k^2  -M^2\,\big)\, {\cal A}_0=0\,.
\label{a0}
\ee
For the longitudinal mode, we turn on both $A_z$ and $A_1$, which
leads to the equation
\be
z^2\,\Big({h}\,{\cal A}'_1\Big)'+ {2k^2 z^3\over M^2-z^2 k^2}\,h \,{\cal A}'_1
+
(z^2 k^2  -M^2)\, {\cal A}_1=0\,,
\label{longi}
\ee
with
\be
{\cal A}_z={k \, z^2\over M^2-z^2 k^2} {\cal A}'_1\,.
\ee
Therefore there are three independent modes for each massive gauge 
field. In order to find the eigenvalue mass scales,  we require
again $a_{\bar{\mu}}$ to zero and impose the horizon boundary condition 
where we discard the logarithmically singular term\cite{Brower:2000rp}. 
Namely suppose
the solution for vector and gravity modes behaves as
\be
{\cal S}_{\rm mode}= C_1\, (1-z)^\beta +C_2 \,(1-z)^\beta \ln\,(1-z) + \cdots
\ee
near horizon region. 
We find that $\beta =0$ for 
$\Phi,\,\, A_1, \, \,A_2,\,\,G_S$  while $\beta=1$ for $A_0,\,\, G_T$. 
Then we choose the boundary condition,
\be
(1-z)\, \Big[\,{{\cal S}_{\rm mode}(z)\over(1-z)^\beta}
\,\Big]' \vert_{z=1}=0\,.
\ee 

\begin{table}[bth]
{
\renewcommand{\arraystretch}{1}
\begin{tabular*}{0.9\textwidth}{@{\extracolsep{\fill}} |c||r|rrr|rr| }
\hline \hline
 \phantom{aaaaaaa} &   \multicolumn{1}{l|}{\phantom{a}spin 0} 
 & \multicolumn{3}{c|}{spin 1}   &\multicolumn{2}{c|}{spin 2} \\
  &       &  \multicolumn{1}{c}{$A_2$}  &  \multicolumn{1}{c}{$A_0$}   
&  \multicolumn{1}{c|}{$A_1$}  &\multicolumn{1}{c}{$G_S$}  
&\multicolumn{1}{c|}{$G_T$} \\
\hline \hline
     &    .699401           &   &        &        &  &   \\
$\Delta=1$ & 2.82963      &        &   &        &        &   \\
     &      5.05667         &   &   &     &        &   \\
\hline
                   &  1.71821      & 1.49111    &  2.72176    &   &    &  \\
$\Delta=2$ &  3.93965  & 3.85854   &    4.99591  &          &   &  \\
                   &   6.17364      & 6.12306  &   7.25013  &        &     &  \\
\hline
                &  2.72176  & 2.56799    &  3.79539    & 1.46635      &2.14858 & 3.60605      \\
$\Delta=3$ &  4.99591   & 4.92662  &6.07826     &   3.93809    & 4.79041        &5.96935  \\
               &  7.25013  &  7.20491   &    8.33987      & 6.22908  &7.11657  & 8.26276   \\
\hline
                     &  3.72304  &    &  &    &   &     \\
$\Delta=4$   &  6.03029 &    &  &    &   &     \\
                     &  8.30414  &    &  &    &   &     \\
\hline
  	              & 4.72364  &  4.62907  &5.85503  & 3.53049   &   &     \\
 $\Delta=5$   & 7.05352   & 6.99910   & 8.17063 & 6.00571   &   &     \\
                      & 9.34436   &  9.30618  & 10.4567 & 8.32808   &   &     \\
\hline
 	              & 5.72397 &    &  &    &   &     \\
 $\Delta=6$   & 8.07028    &    &  &    &   &     \\
                      & 10.3755    &    &  &    &   &     \\
\hline
\end{tabular*}
\caption
    {\small The low lying decay mass scales up to operator
dimension 3 are presented in the unit $4\pi T/3$. 
We also include the mass scales for all the remaining
 SU(4) invariant modes of $\Delta=4,\,5,\,6$.}
\label{tab2} }
\end{table}

We now turn to the case of massless gauge field. The longitudinal 
mode in (\ref{longi}) 
is no longer describing an extra degree and can be discarded
by a gauge fixing conditions, $A_1=0$ and $\partial_\mu 
\sqrt{g} A^\mu=0$,
 while the equations in (\ref{a1}) and (\ref{a0}) are for 
two transverse degrees of the massless gauge field. Nonetheless,
one may evaluate eigenvalues arising from (\ref{longi}) for 
$M^2=0$. We find that the results precisely agree with
those from the transverse mode of (\ref{a0}). In fact one may show 
that the equation (\ref{longi}) follows from (\ref{a0}) by the change 
of the gauge fixing conditions to
$A_0=0$ and $\partial_\mu \sqrt{g} A^\mu=0$. Note also
that the equation (\ref{longi}) for the massless case
agree precisely with the massless scalar equation for 
$\Delta=3$. This explains the full agreement 
between $\Delta=3$ scalar spectra and $\Delta=2$ spectra for
 $A_0$  in Table 2.  

With this setup, we consider the cases of $M^2=0,2,12$
which correspond to the dimensions of current operators
$\Delta =2,3,5$ respectively. The resulting eigenvalues
are presented in Table 2.

Finally we consider the metric perturbation
\be
\delta R_{\mu\nu}+3 \delta g_{\mu\nu}=0\,,
\ee
which is describing the linear  fluctuation of 
the massless AdS$_4$ graviton. Near the boundary $z=0$,
the graviton fluctuation  behaves as
\be
\delta g_{\bar{\mu}\bar{\nu}}=a_{\bar{\mu}\bar{\nu}}(x)\,\, z^{1-\Delta}
+ b_{\bar{\mu}\bar{\nu}}(x)\,\, z^{\Delta-2}+\cdots
\ee
where $\Delta=3$ for our massless metric perturbation.
The field $\delta g_{\bar{\mu}\bar{\nu}}$ is dual to 
the energy momentum tensor operator of the boundary field theory.
Its operator dimension that is protected quantum mechanically
is three as just stated. 
There are two independent physical modes, whose detailed forms
will be identified
below.
We shall use the ansatz
\be
\delta g_{\mu\nu}= {\cal G}_{\mu\nu}(z)\,\, e^{-k x_1}\,.
\ee 
For the tensor mode, we turn on only ${\cal G}_{02}$ component, 
which leads to 
\be
h\,\Big( z^2\,{{\cal G}}'_{02}\Big)'+ \big(\,z^2 k^2  -2 h\,
\big)\, {{\cal G}}_{02}=0\,.
\ee
The treatment of the scalar mode is more involved. We turn on
 ${\cal G}_{00}$, ${\cal G}_{11}$, ${\cal G}_{22}$, ${\cal G}_{zz}$
and ${\cal G}_{1z}$, which turns out to be consistent\cite{Constable:1999gb}. 
The 
Einstein equations imply  that
\be
{\cal G}_{22}=-{1\over h}\,\,{\cal G}_{00}\,, \ \ \ \ \  {\cal G}_{zz}=
{3(1-h)\over h^2 (3+h)}\,\, {\cal G}_{00}\,,
\ee  
and
\be
2k\, {\cal G}_{1z}
=   -{1\over z^2}\big(z^2\,{\cal G}_{11}
\big)'
+{3(1-h)\over z^2\, (3+h)}\,\,\Big({z^2\,{\cal G}_{00}
\over h}\Big)'
+
{18(1-h)\over z\, h^2 (3+h)}\,\,{\cal G}_{00}\,,
\label{1z}
\ee
where ${\cal G}_{11}$ is the gauge freedom. Thus ${\cal G}_{1z}$
can be set to zero by adjusting ${\cal G}_{11}$.
For our purpose,  we choose instead
\be
 {\cal G}_{11}=
{3(1-h)\over h (3+h)}\,\,{\cal G}_{00}\,,
\ee
such that
\be
2k\, {\cal G}_{1z}=
{36(1-h)\over z\, h^2 (3+h)}\,\, {\cal G}_{00}\,.
\ee
This leads to the equation
\be
(z\,h)^2\,\Big( {{\cal G}}''_{00}+
{k^2\over h}\,{\cal G}_{00}
\Big)+
z\, h\,(3-h)\,{\cal G}'_{00}
+\Big(
27+ 9h-51h^2-h^3
\Big)
{(3 - h)\over (3+h)^2}\, {\cal G}_{00}=0\,.
\ee
Further introducing
${\cal S}= 
{\cal G}_{00}/h\,,$
one is led to the equation,
\be
\Big(\, z^2 h\,{\cal S}'\Big)'
+
{z^2 \,k^2}\,{\cal S}+
{4h\over (3+h)^2}\, \Big(
h^2+ 18h-27
\Big)\, {\cal S}=0\,,
\ee
which will be the starting point of our analysis for the scalar graviton 
mode. 
For the length scales, again we require $a_{\bar{\mu}\bar{\nu}}$ to vanish
 and impose the horizon 
boundary condition that the logarithmically singular terms
should be absent. The results are presented in Table 2.

\section{True mass gap and Debye mass}
From the analysis of the last section, one finds that
the lowest value lies in the scalar sector of $\Delta=1$. 
Thus we find  
\be
m_g= 0.699401 \ \ (\,4\pi T/3\,)\,.
\ee
This scalar belongs to the parity even sector so that
the corresponding scalar operator belongs to the $CT$ even 
sector as expected generally for  $CT$ invariant  
theories. The corresponding operator can be constructed using the 
scalars and fermions ($\,\phi^I$, $\psi_J$) of 
the ABJM 
theory, where upper (lower) indices
$I,J=1,2,3,4$ label the SU(4) fundamental 
 (anti-fundamental)
representation.
The chiral primary operator of $\Delta=1$ takes
the form
\be
O_I^J={\rm tr} \,\, \phi^\dagger_I\,\phi^J -
{1\over 4}\,\delta_I^J\,{\rm tr} \,\, \phi^\dagger_K\,\phi^K\, 
\ee
which is traceless.
This corresponds to the (101) representation of 15 components.

For the Debye screening mass,
one needs to
identify the $CT$ eigenvalue of  each operator. 
For the scalar and pseudo scalar, the corresponding operators has
$CT=+1$ and $-1$ respectively. Namely $CT[\Phi^\pm]= \pm 1$ where
the superscript denotes the parity of each mode. 
Similarly one has $CT[A_0^\pm]= \pm 1$,
$CT[A_1^\pm]=CT[A_2^\pm]= \mp 1$ for the vector operators
and $CT[G_S]=-CT[G_T]=+1$ for the energy momentum tensor.
Therefore we conclude that
\be
m_D= 1.71821 \ \ (\,4\pi T/3\,)\,,
\ee
arising from the $\Delta=2$, parity-odd scalar mode.
The corresponding operators can be constructed as follows.
Let us first consider the chiral primary operator
\be
O_{IJ}^{KL}={\rm tr} \,\, \phi^\dagger_{\left(I\right.}\,\phi^{
\left(K\right.} \,
\phi^\dagger_{\left.J\right)}\,\phi^{\left.L\right)} 
-({\rm trace \ part})
\ee
which has total  84 independent components. This corresponds to
 $\Delta=2$ operator  of the $(202)$ representation, which is
 not the desired one
since we are looking for the operator of 
(101) representation.
The only remaining possibility 
is\footnote{We follow the spinor convention in Ref.~\cite{Bak:2008cp}.} 
\be
\tilde{O}_I^J={\rm tr} \,\,\psi_I \, \psi^{\dagger\,J}-
{1\over 4}\,\delta_I^J\,{\rm tr} \,\,\psi_K\, \psi^{\dagger\,K}\,, 
\ee
which may be obtained from $O_I^J$ by acting supercharges 
twice. 
This has all  the desired properties including the parity oddness
and, with its help, one may probe the Debye screening mass.

 Final comments are in order. 
The ratio $m_D/m_g$ at strong coupling limit takes the value
$2.45669$ and its comparison with those of  other
Chern-Simons-matter theories will be of interest. 
In Ref.~\cite{Smedback:2010ji},
the thermal scalar mass scale is obtained 
in the weak coupling regime of the ABJM field 
theory. 
Relating this scalar mass scale 
to our mass scales calls for a  closer look.
Once introducing the scalar mass terms, one can also find
the gauge self-energy part $\Pi_{00}(0)$ following the computation 
in \cite{Pisarski:1986gq}. Due to the difference in the form of the 
classical (Chern-Simons) kinetic term  from that of 
the ${\cal N}=4$ SYM theory, the identification of the decaying mass
scale requires an additional  work compared to that of 
the Debye mass of the ${\cal N}=4$ 
SYM theory\cite{Kim:1999sg}.
 After surveying
possible mass scales, one can identify the weak coupling behaviors
of the Debye mass and, perhaps, the true mass gap if it does not 
include any non perturbative effects. We view this problem very 
interesting  but it requires a separate  study.

\section*{Acknowledgement}
KB would like to thank M.M. Sheikh-Jabbari for valuable discussions.
This work was supported in part by KRF R01-2008-000-10656-0,
SRC-cquest-R11-2005-021.


\end{document}